# Traffic Load Uniformity in Optical Packet Switched Networks

Harald Øverby
*Norwegian University of Science and Technology*

*Abstract*— **A crucial issue in Optical Packet Switched (OPS) networks is packet loss due to contentions. In this paper we study how traffic load uniformity influences the packet loss in a single OPS node. Traffic load uniformity is a measure of the load asymmetry a specific output link in an OPS node receives from its input sources. We develop analytical models based on the Engset traffic model. As a major contribution of this paper, we show that an asymmetric traffic pattern results in less packet loss compared to a symmetric traffic pattern for a single output link in an OPS node.**

## I. INTRODUCTION

Optical Packet Switching (OPS) is a promising architecture for the future core Internet, taking advantage of optical processing and data combined with statistical multiplexing for excellent resource utilization [2,10]. A crucial issue in OPS networks is packet loss due to contentions [5,6,7,12]. Contentions occur when two or more packets are aligned toward the same output link at the same time, leading to an increased Packet Loss Ratio (PLR).

The number of contentions occurring in an OPS network is heavily influenced by the traffic pattern in the network, including the burstiness of the traffic and the network topology [11]. For instance, it has been shown that bursty traffic leads to an increased PLR in OPS [11]. An important aspect of the traffic pattern is traffic load uniformity [4,13], defined as the load distribution from a set of sources on a given output link. The main goal in this paper is to quantify how the traffic load uniformity influences the PLR in a single OPS node.

Fig. 1 provides a simple illustration. Consider two input channels (IC) in an optical packet switch direct traffic to a tagged output link (TOL). First, consider the case with symmetric traffic, with a load $A_1=A_2=0.4$ Erlang on each input channel $IC_1$ and $IC_2$. The resulting PLR is, according to the Engset loss formula [9,14] (traffic congestion), $E_2(0.4+0.4)=28.6$ %. Second, consider the case with asymmetric traffic, with loads $A_1=0.7$ and $A_2=0.1$. The total load is the same as in the symmetric case, however, the PLR is now only maximum $0.1/(0.1+0.7)=12.5$ %, assuming that all packets from $IC_1$ are transmitted and packets from $IC_2$ are dropped in the case of contention. Further reducing the traffic on $IC_2$ while holding the total traffic load the same, will eventually lead to a PLR=0, as only $IC_1$ is transmitting traffic.

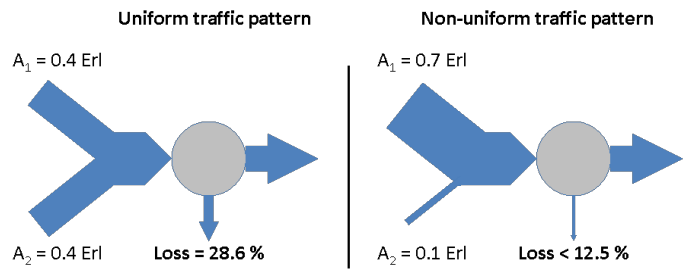

Fig. 1. A simplified example illustrating the impact of traffic load uniformity on the PLR. A total load of 0.8 Erl is offered from two IC. When the load is uniformly distributed over the two IC, the resulting PLR is 28.6 %. When the load is non-uniformly distributed, the resulting PLR is at maximum 12.5 %.

The aim in this paper is to quantify this asymmetry using the Traffic Uniformity Index (TUI) and study how the TUI influences the PLR in a single OPS node. We will also develop analytical models capturing the TUI influence in the PLR in OPS nodes using the Engset traffic model. As to our knowledge, this issue has not been addressed in the context of OPS, and will contribute to an increased understanding of the theoretical fundaments of OPS networks.

## II. RELATED WORKS

Related works includes teletraffic models for optical packet switched networks, which have been extensively studied during the last years. General analytical models based on the Erlang Traffic model for slotted [15] and asynchronous [16] OPS have been outlined. These models have proven to be accurate with respect to estimate the PLR for both OPS and Optical Burst Switched (OBS) networks. However, the Erlang traffic model cannot capture the effects of asymmetric traffic, as it assumes that packets arrive to a TOL according to a single poisson process with rate λ. The Engset traffic model [9,14] models the individual input channels as separate on/off processes, and has been proven to be even more accurate than models based on the Erlang traffic model, at the expense of an increased computational complexity. However, the works in [9,14] assume a symmetric traffic pattern, and thus ignore the impact of non-uniform traffic.

A study of Optical Burst Switched (OBS) networks under a non-uniform traffic distribution have been considered in [8]. However, the paper [8] considers the effects of non-uniform traffic on the network level, where this paper consider non-uniform traffic on the node level, i.e. for an OPS node. The effects of non-uniform traffic have also been studied for an optical slotted ring network [13]. A quantifiable measure of the traffic uniformity, suitable for network level studies, was also proposed in [13]. This measure has been used as a basis for the TUI measure presented in this paper. The authors of [3] show that asymmetric or non-uniform traffic better estimates the real Internet traffic. This motivates the topic for this paper, as asymmetric traffic patterns should be studied for performance studies for OPS.

## III. SYSTEM MODEL

We consider a tagged output link (TOL) in an optical packet switch consisting of F input/output fibres and W wavelength channels per fibre, as seen in Fig. 2. The TOL receives packets from a number M input channels, denoted $IC_i$ ($1 \leq i \leq M$). Note that M is a general parameter and may be interpreted as e.g. the number of input fibres, M=F (in the case of an OPS without wavelength conversion), or the number of input wavelengths, M=FW (in the case of an OPS with full-range wavelength converters). Each IC is modelled according to an on/off source. The optical packet switch does not have buffering capabilities for contention resolution, however, the switch may have wavelength conversion capabilities, depending on the switch configuration, as detailed in Section 4.

Packets arrive on $IC_i$ from adjacent OPS nodes according to an on/off arrival process with normalized load $A_i$, arrival intensity $\lambda_i$, and departure intensity $\mu$. Note that the arrival intensity is unique for each IC, reflecting the possible asymmetry of the input traffic. However, the departure intensity is common for all packets, reflecting that the packet length distribution is identical and not dependent on the IC. The arrival intensity in $IC_i$ is given by $\lambda_i = A_i \mu / (1 - A_i)$. Packets that cannot be accommodated on an available wavelength is dropped and contributes to an increased Packet Loss Ratio (PLR). Furthermore, for simplicity and without loosing generality, we normalize the departure intensity to $\mu=1$.

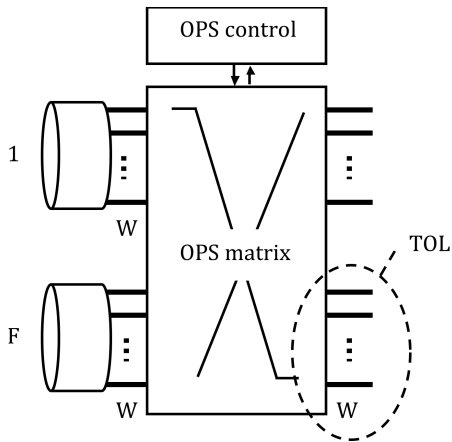

Fig. 2. The OPS node under study with M=FW input channels and W output channels at the Tagged Output Link (TOL).

We quantify the traffic load uniformity using the traffic uniformity index (TUI). We consider a TOL receiving traffic from M input channels (IC). As seen in section 2.1, the normalized load on $IC_i$ is $A_i$. The TUI of the TOL is then defined as:

$$TUI = \frac{\left(\sum_{i=1}^{M} A_i\right)^2}{M \sum_{i=1}^{M} A_i^2} \quad \left(\frac{1}{M} \leq TUI \leq 1\right) \quad (1)$$

The TUI produces a number in the range [1/M,1]. A TUI value of 1 refers to perfect symmetry, i.e. traffic to the TOL is evenly distributed among all IC. A TUI value of 1/M refers to completely asymmetry, i.e. all traffic arriving at the TOL is routed from a single IC.

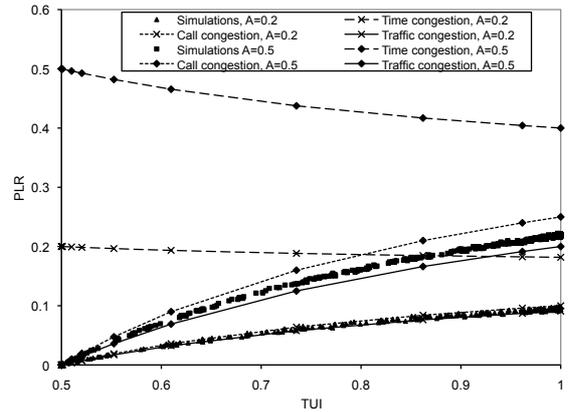

Fig. 3. The PLR as a function of the TUI for different values of the system load and blocking metrics for configuration 1. The analysis is based on the Engset LCC. M=2 and W=1.

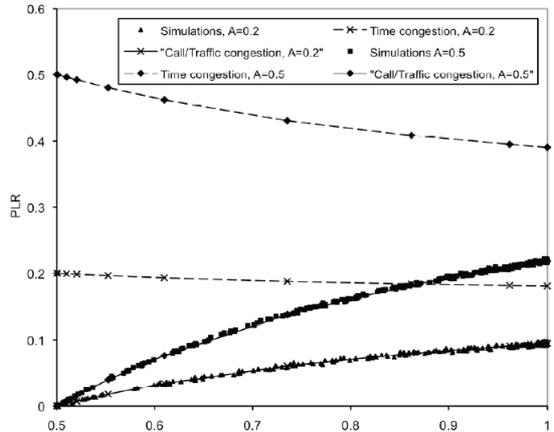

Fig. 4. The PLR as a function of the TUI for different values of the system load and blocking metrics for configuration 1. The analysis is based on the Engset OFL. M=2 and W=1.

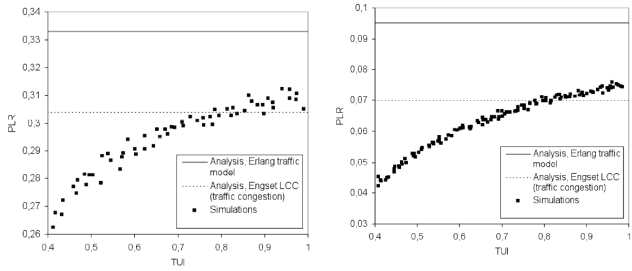

Fig. 5. The PLR as a function of the TUI for M=8, W=1, A=0.5 (a). The PLR as a function of the TUI for M=16, W=4, A=0.5 (b).

## IV. RESULTS

We report results based on analytical and simulation studies. Figs. 3 and 4 illustrate the results obtained from the analysis and the simulations for configuration 1. The analytical results are obtained using the Engset LCC and the Engset OFL [14], respectively. Regarding the call- and traffic congestion blocking metrics, and the simulations, we observe that the PLR decreases as the TUI decreases, indicating that an increased asymmetry of the input traffic results in a decreased PLR. For the time congestion blocking metric we observe that the PLR slightly decreases as the TUI increases. This indicates that the time congestion is unsuitable to calculate the PLR in this scenario. On the other hand, both the call- and traffic congestion blocking metrics match well with the results obtained from the simulations. We observe that when TUI approaches 0.5, the PLR approaches 0.

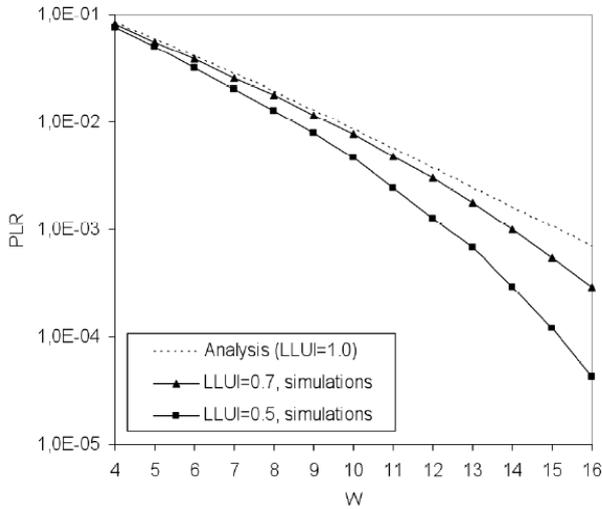

Fig. 6. The PLR as a function of W for A=0.5 and M=32.

Fig. 5 shows the PLR as a function of the TUI. We observe that the PLR decreases as the TUI decreases. On both figures we see that the traditional Engset traffic model estimates the PLR when TUI=1.0 accurately, however, it fails to estimate the PLR accurately when TUI<0.8. Hence, both the Engset LCC traffic model, and the Erlang traffic model, in particular, are not suitable to model asymmetric traffic in OPS nodes.

Fig. 6 shows the PLR as a function of W. Ten independent simulations runs have been performed in order to obtain 95 % confidence limits. We observe that the traditional Engset traffic model estimates the PLR accurately for W<8. For large values of W, the traditional Engset traffic model fails to accurately estimate the PLR, especially for low values of the TUI.

## V. CONCLUSIONS

In this paper we presented how asymmetric traffic influences the packet loss ratio in an optical packet switch. We have quantified the degree of traffic asymmetry using the traffic uniformity index. Our findings show that increasing the asymmetry of the input traffic leads to a reduced packet loss ratio for OPS.